\documentclass[twocolumn, switch]{article} 

\usepackage{preprint}

\usepackage{amsmath, amsthm, amssymb, amsfonts, tabularray, rotating, enumitem, caption}

\usepackage[numbers,square]{natbib}
\usepackage[utf8]{inputenc}	
\usepackage[T1]{fontenc}	
\usepackage{xcolor}		
\usepackage[colorlinks = true,
            linkcolor = purple,
            urlcolor  = blue,
            citecolor = cyan,
            anchorcolor = black]{hyperref}	
\usepackage{microtype}		
\usepackage{float}			
\usepackage{multicol}		

\usepackage{titlesec}
\titlespacing\section{0pt}{12pt plus 3pt minus 3pt}{1pt plus 1pt minus 1pt}
\titlespacing\subsection{0pt}{10pt plus 3pt minus 3pt}{1pt plus 1pt minus 1pt}
\titlespacing\subsubsection{0pt}{8pt plus 3pt minus 3pt}{1pt plus 1pt minus 1pt}

\usepackage{tikz,xcolor,hyperref}

\definecolor{lime}{HTML}{A6CE39}
\DeclareRobustCommand{\orcidicon}{
	\begin{tikzpicture}
	\draw[lime, fill=lime] (0,0) 
	circle [radius=0.16] 
	node[white] {{\fontfamily{qag}\selectfont \tiny ID}};
	\draw[white, fill=white] (-0.0625,0.095) 
	circle [radius=0.007];
	\end{tikzpicture}
	\hspace{-2mm}
}
\foreach \x in {A, ..., Z}{\expandafter\xdef\csname orcid\x\endcsname{\noexpand\href{https://orcid.org/\csname orcidauthor\x\endcsname}
			{\noexpand\orcidicon}}
}

\title{Market-Derived Financial Sentiment Analysis: Context-Aware Language Models for Crypto Forecasting}

\usepackage{authblk}

\author[1]{Hamid Moradi-Kamali\orcidA{}}
\author[1]{Mohammad-Hossein Rajabi-Ghozlou\orcidB{}}
\author[1]{Mahdi Ghazavi\orcidC{}}
\author[1]{Ali Soltani\orcidD{}}
\author[1]{Amirreza Sattarzadeh\orcidE{}}
\author[1]{Reza Entezari-Maleki\orcidF{}}

\affil[1]{School of Computer Engineering, Iran University of Science and Technology, Tehran, Iran}

\begin{document}

\twocolumn[ 
  \begin{@twocolumnfalse} 
  
\maketitle

\begin{abstract}
Financial Sentiment Analysis (FSA) traditionally relies on human-annotated sentiment labels to infer investor sentiment and forecast market movements. However, inferring the potential market impact of words based on their human-perceived intentions is inherently challenging. We hypothesize that the historical market reactions to words, offer a more reliable indicator of their potential impact on markets than subjective sentiment interpretations by human annotators. To test this hypothesis, a market-derived labeling approach is proposed to assign tweet labels based on ensuing short-term price trends, enabling the language model to capture the relationship between textual signals and market dynamics directly. A domain-specific language model was fine-tuned on these labels, achieving up to an 11\% improvement in short-term trend prediction accuracy over traditional sentiment-based benchmarks. Moreover, by incorporating market and temporal context through prompt-tuning, the proposed context-aware language model demonstrated an accuracy of 89.6\% on a curated dataset of 227 impactful Bitcoin-related news events with significant market impacts. Aggregating daily tweet predictions into trading signals, our method outperformed traditional fusion models (which combine sentiment-based and price-based predictions). It challenged the assumption that sentiment-based signals are inferior to price-based predictions in forecasting market movements. Backtesting these signals across three distinct market regimes yielded robust Sharpe ratios of up to 5.07 in trending markets and 3.73 in neutral markets. Our findings demonstrate that language models can serve as effective short-term market predictors. This paradigm shift underscores the untapped capabilities of language models in financial decision-making and opens new avenues for market prediction applications.
\end{abstract}
\vspace{0.35cm}

  \end{@twocolumnfalse} 
] 

\section{Introduction}
Today, a single social media post from a public figure can trigger billion-dollar fluctuations in global markets within minutes. As markets evolve, they are increasingly shaped by the collective psychology and behavior of investors. Investor sentiment, often reflected in social media posts on platforms such as X (formerly Twitter), serves as a proxy for market expectations \cite{elon_ante_2023}. FSA quantifies these sentiments, typically classified as \textit{Bullish}, \textit{Bearish}, or \textit{Neutral}, to forecast market movements. However, significant challenges remain in translating these classifications into actionable trading insights. First, sentiment classifications often lack precision and explainability. For instance, a \textit{Bullish} label might indicate anything from a modest 2\% price increase over five days to a substantial 10\% rise over a month, making it difficult to establish clear trading strategies based on such vague signals \cite{du2024financial}. Second, traditional FSA relies on human-annotated sentiment labels, which embed subjective interpretations of a text’s market impact. These preconceived notions often diverge from historical market reactions, introducing a gap between sentiment perception and measurable outcomes. Finally, financial tweets differ significantly from traditional textual data due to their brevity, domain-specific jargon, and references to real-time events. Ignoring this context diminishes the relevance of sentiment analysis for predicting market trends \cite{man2019financial}.
To address these limitations, we propose a novel approach to FSA that focuses on measurable market reactions rather than inferred intentions. Instead of relying on subjective, human-annotated sentiment labels, market-derived annotation is adopted, which directly ties labels to subsequent market movements. This approach leverages financial markets' well-defined metrics to replace manual annotations with precise, automated labels \cite{du2024financial, advances_deprado_2018}. By grounding sentiment in historical market responses, a closer alignment between sentiment analysis and actionable trading strategies is ensured. Additionally, contextual market data is incorporated into the analysis using a prompt engineering technique which enables language models to integrate key market indicators, such as price trends and market volatility directly into textual analysis \cite{leveraging_ahmed_2024}. By embedding this context into tweets, our method transforms FSA into a unified framework that synthesizes sentiment analysis and market prediction into one. Unlike traditional multi-model approaches, our framework eliminates the need for separate price prediction models and fusion layers, offering a language model approach to the problem of FSA. 

\begin{figure*}[t]
    \centering
    \includegraphics[width=0.8\linewidth]{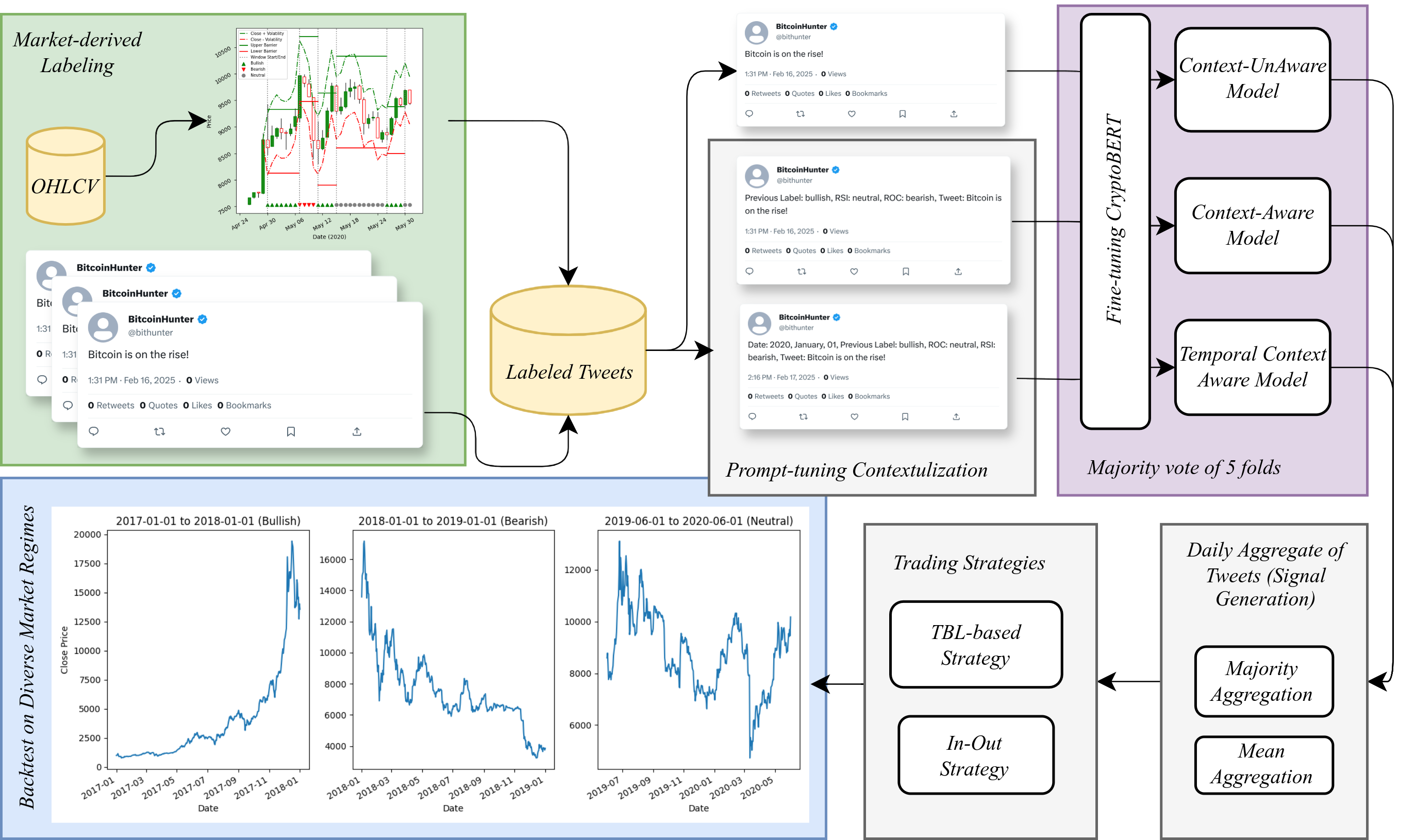}
    \caption{Overall scheme of the proposed approach}
    \label{fig:scheme}
\end{figure*}

This study introduces a new perspective on sentiment analysis in finance, demonstrating that language models can be effective short-term Bitcoin market predictors. This contrasts starkly with previous approaches where language models, relying on human-annotated sentiment or lacking market context, have shown limited predictive capability on their own. Our work thus calls to revisit the assumptions underlying traditional FSA, highlighting the untapped potential of language models for short-term market prediction when properly fine-tuned. Specifically, our main contributions are as follows:
\begin{itemize}[leftmargin=5mm]
\item Market-derived Labeling: We replace subjective sentiment labels with market-derived labels that directly reflect ensuing price movements, enabling the model to learn from actual market behavior rather than inferred sentiment.
\item Contextualized Prompt-tuning: By embedding key market information into textual prompts, our approach expands the feature space of language models, providing the model with the context required for precise market predictions.
\item Trading Strategy Evaluation: Model predictions are translated into trading signals and rigorously assessed through backtesting over three years of Bitcoin market data, covering bullish, bearish, and neutral market regimes.
\end{itemize}
This simple end-to-end language-model approach, shown in Figure~\ref{fig:scheme}, offers a practical and robust solution to FSA, demonstrating the potential of such an approach by effectively unifying textual and market data within a single model.

The remaining part of this paper is organized as follows. Section \ref{sec:background} provides a background to the labeling method, and language models used in the rest of the paper. Section \ref{sec:related_works} presents the related works that inspired this study. Section \ref{sec:proposed} explains the labeling process, proposed language models, prompt tuning process, proposed signal generation methods, and backtesting strategies. Section \ref{sec:results} presents and interprets the experimental results on tweet classification, signal generation, and backtesting of our proposed methods, along with their comparison to the benchmarks. Section \ref{sec:discussion} involves this study's limitations and implications of this study, and Section \ref{sec:conclusion} concludes the study's highlights and suggests multiple future directions for the readers.

\section{Background}
\label{sec:background}
This section presents the challenges of market-derived labeling and Triple Barrier Labeling (TBL) \cite{advances_deprado_2018} as the basis for our proposed solution. We also introduce CryptoBERT and FinBERT as the base domain-specific language models for the fine-tuning process.
\subsection{Triple Barrier Labeling}
Labeling short-term market trends as \textit{Bullish}, \textit{Bearish}, or \textit{Neutral} requires a method that is precise, adaptable, and aligned with trading applications. In this context, short-term trends represent the overall price direction of an asset over a defined timeframe (typically 8 to 15 days, as supported by studies of post behavior on X \cite{valle2022does}). TBL provides a robust framework for this purpose by incorporating three distinct barriers that capture market behavior dynamically:  
\begin{enumerate}[leftmargin=5mm]
    \item Upper Barrier: A profit-taking price threshold that, when reached, labels the trend window as \textit{Bullish}.
    \item Lower Barrier: A stop-loss price threshold that, when breached, labels the trend window as Bearish.
    \item Vertical Barrier: A temporal limit that marks the expiration of an observation window, ensuring trends are assessed within a fixed timeframe.
\end{enumerate}

Unlike traditional fixed-time horizon methods, TBL is a path-dependent labeling strategy, meaning it evaluates the entire price trajectory within the observation window. Labels are determined by the first barrier touched during this period, with dynamic adjustments based on market volatility \cite{advances_deprado_2018}. For instance, the horizontal barriers (upper and lower) are scaled by a volatility estimate, ensuring that they are sensitive to the asset’s price fluctuations. The vertical barrier ensures that observations remain time-bounded, preventing positions from being unrealistically held indefinitely. TBL addresses several key challenges in financial labeling. It provides an intuitive and explainable labeling mechanism that aligns with real-world trading practices \cite{bounid2022advanced}. By dynamically adjusting barriers based on market volatility, it remains robust across different market regimes. Its path dependence ensures that labels reflect the actual price trajectory rather than arbitrary points within a fixed timeframe, reducing look-ahead bias. For example, in a volatile Bitcoin market, TBL adjusts the horizontal barriers to accommodate rapid price swings, ensuring that labels remain consistent with actionable trading opportunities. This adaptability makes TBL highly suitable for live trading and machine-learning applications \cite{fu2024enhanced}. Moreover, by combining profit-taking, stop-loss, and temporal constraints, TBL provides a comprehensive framework that captures market dynamics effectively while mitigating overfitting risks.

\section{Related Works}
\label{sec:related_works}
Sentiment analysis in financial markets emerged from the growing recognition that market psychology plays a crucial role in market movements \cite{adaptive_lo_2004}. As Natural Language Processing (NLP) techniques advanced, they enabled the extraction of sentiments and emotions from large volumes of unstructured text, providing a means to quantify market psychology \cite{twitter_bollen_2011}. Social media platforms, particularly Twitter, became valuable sources of real-time insights into investor sentiment, which could be used to predict financial market behavior. Sentiment analysis, or opinion mining, had already been established as a method within NLP for extracting emotional responses \cite{murfi2024bert}, making them well-suited for quantifying investor psychology and applying these signals to financial markets. The power of sentiment analysis in forecasting market trends, volatility, volume, and risk contributed to the birth of FSA \cite{man2019financial, du2024financial}. The primary goal of FSA has been to utilize NLP methods to capture investor sentiment and psychology, providing valuable insights for predicting market movements \cite{du2024financial}. Over time, the field has evolved, and various methods have been developed to refine sentiment classification, improve prediction accuracy, and integrate more nuanced factors such as market context. In the subsequent sections, we will explore the current state of the art in FSA, examine existing gaps, and highlight how this study aims to address these challenges.

\subsection{Financial Sentiment Analysis Scope}
FSA spans a wide range of methods, including lexicon-based approaches \cite{du2023finsenticnet}, word representations \cite{cryptocurrency_lamon_2017}, and pre-trained language models \cite{teng2025bert, huang2023finbert, sentiment_kulakowski_2023, fingpt_yang_2023, fineas_gutirrezfandio_2021}. It also draws on various data sources such as news headlines \cite{cryptocurrency_lamon_2017}, social media posts \cite{singh2020sentiment}, and financial documents \cite{man2019financial}, with applications in volatility prediction, price prediction, and trend forecasting \cite{du2024financial}. The core goal of FSA is to quantify investor sentiment from textual data to inform decision-making in financial markets. In this study, we focus on predicting Bitcoin’s short-term market movements by quantifying sentiment from Bitcoin-related tweets. This task has been addressed by various studies, but we consider PreBit \cite{prebit_zou_2023} to be the state-of-the-art in this area. Unlike many other studies, PreBit advances the field by passing the actual tweet representations, rather than their sentiment class labels, to the trend prediction model. This allows the model to learn patterns between the words in the tweets and the market movements they ought to predict. By retaining the tweet’s representation, the model can identify nuanced relationships between specific words and their effect on market behavior, bridging the gap between textual sentiment and market outcomes.

Despite advancements like these, FSA models are still often used in conjunction with price-based prediction models because they do not perform adequately on their \cite{man2019financial}. This limitation is not necessarily due to the inferiority of textual data compared to price-based features. Even when domain-specific, pre-trained word embeddings are used, the issue lies in how we still assign sentimental value to words based on subjective perceptions, rather than allowing the market’s reaction to define these relationships. This leads to our core hypothesis: market-driven labels, learned from actual market movements, can better inform sentiment analysis models. By letting the market itself teach the model the significance of words, we aim to bridge the gap between textual content and market reactions, moving away from subjective interpretations. Our hypothesis also points to a second major challenge: contextualization. The context in which a tweet is shared—such as market conditions, sentiment trends, or external factors—can be more influential than the content itself. Previous research supports this idea, showing that expanded feature spaces, including factors like author credibility \cite{elon_ante_2023, GARCIAMENDEZ2024124515, ask_akbiyik_2021}, post volume \cite{bitcoin_critien_2022}, and extracted topics of discussion \cite{hu2021stock}, significantly improve the performance of FSA models. In the next section, we discuss the latest developments in NLP and FSA that have inspired our research and highlight how our approach builds on these advances.

\subsection{Innovations in NLP, and The FSA Horizon}
Large Language Models (LLMs) have increasingly been leveraged for financial market analysis, with prominent examples such as FinBERT \cite{huang2023finbert}, CryptoBERT \cite{sentiment_kulakowski_2023}, and FinEAS \cite{fineas_gutirrezfandio_2021}. These models have set a strong foundation for improving FSA tasks. FinBERT, for instance, adapts to the finance domain by effectively incorporating contextual information from financial texts, outperforming traditional machine learning algorithms in sentiment classification. Similarly, FinEAS, building upon Sentence-BERT, generates high-quality sentence embeddings that substantially enhance FSA task performance. CryptoBERT, which serves as the foundation for our study, is fine-tuned on cryptocurrency-related social media data, leveraging the BERTweet architecture and domain-specific training to accurately analyze crypto-related tweets. On the other hand, FinGPT \cite{fingpt_yang_2023} represents a breakthrough as a general-purpose financial advisor, demonstrating how LLMs can synthesize market data and provide accessible insights for developing financial models. In our work, we build upon CryptoBERT by incorporating market and temporal context into tweets, enabling the model to better detect patterns and improve prediction accuracy, ultimately offering more actionable insights for cryptocurrency investors. Our study is further inspired by research that examined the temporal window of tweet propagation during events like the COVID-19 and H1N1 pandemics, which suggested that the most impactful tweet propagation occurs within a window of 8 to 15 days \cite{valle2022does, whether_aysan_2023}. This finding informed our decision to focus on short-term trend prediction in the market, aligning with the intuition that market psychology (driven by fear and greed) manifests strongly in short-term fluctuations.

A key innovation in improving text classification tasks like FSA has been prompt engineering \cite{promptlearning_zhu_2022}. Specifically, studies such as Dhar et al. \cite{analysis_dhar_2023} have demonstrated the transformative impact of prompt engineering on financial decision-making, enhancing operational efficiency, risk management, and customer service. Ahmed et al. \cite{leveraging_ahmed_2024} further explored prompt engineering in FSA, showing how carefully designed prompts can significantly boost model performance. In our research, we use prompt fine-tuning to incorporate both temporal and market context, enhancing the model's capacity to identify patterns and improve the quality of insights for financial forecasting. We were also inspired by Dhingra et al. \cite{timeaware_dhingra_2021}, who proposed jointly modeling text with timestamps to improve temporal knowledge. Additionally, Agarwal et al. \cite{natural_xing_2018} demonstrated the benefits of temporal domain adaptation in improving model predictions. By integrating both temporal and market information into the tweet context, we enhance our model's ability to accurately predict market movements and generate actionable insights.

Despite the current limitations and challenges in FSA, the future of this field holds tremendous promise. While there are still gaps to bridge, the continuous advancements in NLP, particularly through LLMs and innovative approaches like prompt engineering, offer a clear path forward. By leveraging market-driven labels, enhancing models with contextual and temporal information, and incorporating new data sources, we can address existing shortcomings and unlock the full potential of sentiment analysis in financial forecasting. As more studies explore these avenues and integrate market-specific context, the scope for FSA to play a central role in predicting market trends, managing risk, and enhancing decision-making in finance continues to expand. With these ongoing innovations, the future of FSA is indeed bright, offering exciting opportunities for more accurate and actionable financial predictions.

\section{Proposed Method}
\label{sec:proposed}
This section presents our proposed method, starting with a description of the datasets and a novel trend-based labeling approach inspired by the TBL method. We then expand the language model’s feature space through prompt tuning, enabling it to capture complex interrelations among textual content, market indicators, and temporal dynamics. We introduce two trading signal generation methods to predict the direction and magnitude of trading positions. You can find all the implementation details in a public repository\footnote{github.com/hamidm21/Revisit\_FSA}. Finally, we introduce our proposed backtesting strategies and evaluation metrics.

\subsection{Data Collection}
We curated datasets comprising daily tweets, technical indicators, and significant Bitcoin events to support market trend prediction using sentiment and technical indicators. Each dataset serves a distinct purpose: tweets are used for model training and evaluation; technical indicators are incorporated as price-based features in classification models and as market context during the prompt-tuning process. also, a dataset of historical events is leveraged to benchmark model performance on key dates, particularly during periods of high market volatility or impactful news cycles. Here we present these datasets in more detail:

\begin{itemize}[leftmargin=5mm]
    \item Bitcoin Historical Events: This dataset contains 227 significant Bitcoin market events from 2009 to 2024, focusing on benchmarking model performance during influential periods between 2015 and 2023. Accessible from Kaggle\footnote{kaggle.com/datasets/hamidmoradi21/bitcoin-historical-events}, the dataset catalogs pivotal news events such as regulatory announcements, major hacks, and institutional adoptions that historically triggered substantial price volatility. By including these events, the dataset enables performance evaluation during market conditions that often require rapid and nuanced sentiment analysis, providing insights into how well the models respond to real-world, high-stakes scenarios.
    \item Tweets: Tweet data from two sources, spanning 2015 to early 2023, was utilized. The first dataset, covering from January 2015 to February 2021, was obtained from PreBit \cite{prebit_zou_2023}, and the second dataset, sourced from Kaggle\footnote{kaggle.com/datasets/kaushiksuresh147/bitcoin-tweets}, extends from February 2021 to March 2023. For each day, tweets were combined and preprocessed through steps including text normalization (lowercasing, removal of URLs, user IDs, and punctuation), extraction of emojis and hashtags as sentiment indicators, and removal of promotional and advertisement content to reduce noise. Lemmatization is also applied to unify terms and improve generalization.

    \item Technical Indicators: The Relative Strength Index (RSI) and Rate of Change (ROC) are calculated from OHLCV\footnote{Open, High, Low, Close, and Volume} data and are used as technical indicators for prompt tuning. Thresholds of 30 and 70 are set for RSI to signal oversold and overbought conditions. ROC thresholds are dynamically adjusted based on the standard deviation of eight-day returns to balance the dataset and align with RSI levels, providing more robust trend insights and identifying potential reversals. The same OHLCV data, spanning from January 2015 to March 2023, was used to produce the daily labels.
\end{itemize}
Together, these datasets form a multi-faceted basis for trend prediction, enhancing accuracy and generalization across varying market conditions.

\subsection{Market-derived Labeling}
Most FSA studies rely on human-annotated sentiment labels to train language models for classifying financial tweets. These sentiment classifications are then used to predict next-day market movements, typically using fixed thresholds for bullish and bearish labels based on the next day’s ROC \cite{john2024cryptocurrency, du2024financial}. However, our approach takes a fundamentally different path by directly using market-derived labels to classify tweets, bypassing the subjectivity of human annotations. While the influence of social media sentiment on financial markets often extends up to 15 days, many studies oversimplify this relationship by relying on single-day thresholds \cite{valle2022does}. Such methods fail to capture the prolonged effects of sentiment and ignore real-world trading practices, where trends span multiple days. Effective labeling must account for these dynamics and incorporate mechanisms like take-profit, stop-loss, and vertical barriers to manage risk and adapt to market volatility \cite{bounid2022advanced}. To address these challenges, we propose a trend-based labeling method inspired by TBL. This approach integrates historical volatility and market-defined barriers, creating explainable and actionable labels that establish a direct connection between textual inputs and market trends. The method dynamically adjusts upper and lower barriers based on market volatility, as detailed below:

\begin{itemize}[leftmargin=5mm]
    \item Historical Volatility Estimation:  
    The historical volatility of log returns is estimated using the Exponentially Weighted Moving Average (EWMA) method. Log returns are calculated as shown in Eq.~(\ref{eq:log_ret}):
    \begin{align}
    \label{eq:log_ret}
    r_t = \ln\left(\frac{P_t}{P_{t-1}}\right)
    \end{align}
    where \( P_t \) represents the closing price at time \( t \). Log returns quantify the relative change in price between consecutive time steps and serve as the basis for volatility estimation. Using the log returns, the EWMA method calculates their exponentially weighted standard deviation. The formula for EWMA volatility is given in Eq.~(\ref{eq:ewma_vol}):
    \begin{align}
    \label{eq:ewma_vol}
    \sigma_t = \sqrt{\frac{\alpha}{1 - (1 - \alpha)^\tau} \sum_{i=1}^{N} (1-\alpha)^{i-1} (r_{t-i})^2}
    \end{align}
    where:
    \begin{itemize}[leftmargin=5mm]
        \item \( \sigma_t \) represents the EWMA volatility, defined as the standard deviation of the exponentially weighted log returns.
        \item \( \alpha = \frac{2}{\tau + 1} \) is the smoothing factor, controlling the decay of weights.
        \item \( \tau \) is the window length (set to 30 days in this study).
        \item \( N \) is the maximum number of terms in the dataset.
    \end{itemize}

    The EWMA volatility emphasizes recent price dynamics by assigning exponentially decreasing weights to older log returns, ensuring that more recent movements have a greater influence on the volatility estimation.

    \item Barrier Calculation and Label Assignment:  
Upper and lower barriers are calculated using the historical volatility (\( \sigma_t \)) derived in Eq.~(\ref{eq:ewma_vol}) as a coefficient to the barrier factors, enabling forward-looking estimations without introducing look-ahead bias. The upper and lower barriers for a given time \( t \) are calculated as shown in Eq.~(\ref{eq:upper}) and Eq.~(\ref{eq:lower}), respectively:
\begin{align}
\text{U}_t &= P_t + (P_t \times \sigma_t \times F_\text{u}) \label{eq:upper} \\
\text{L}_t &= P_t - (P_t \times \sigma_t \times F_\text{l}) \label{eq:lower}
\end{align}
where:
\begin{itemize}[leftmargin=5mm]
    \item \( P_t \) is the closing price at time \( t \),
    \item \( \sigma_t \) is the historical volatility at time \( t \), calculated using Eq.~(\ref{eq:ewma_vol}),
    \item \( F_\text{u} \) and \( F_\text{l} \) are the upper and lower barrier factors, optimized based on market dynamics.
\end{itemize}

A vertical barrier (\( \text{V}_t \)) is defined as the maximum allowable time horizon for labeling, ranging from 8 to 15 days. Label assignment is determined based on the following rules:
\begin{itemize}[leftmargin=5mm]
    \item A \textit{Bullish} label is assigned if \( P_t \geq \text{U}_t \) before reaching \( \text{V}_t \).
    \item A \textit{Bearish} label is assigned if \( P_t \leq \text{L}_t \) before reaching \( \text{V}_t \).
    \item A \textit{Neutral} label is assigned if neither \( \text{U}_t \) nor \( \text{L}_t \) is crossed by the time \( \text{V}_t \) is reached.
\end{itemize}

Once a barrier (\( \text{U}_t \) or \( \text{L}_t \)) is touched, the labeling window closes, and a new window is initiated with updated \( \text{U}_t \), \( \text{L}_t \), and \( \text{V}_t \) values recalculated from the historical volatility at that time. This approach ensures the method remains forward-looking and avoids look-ahead bias, as each decision relies solely on information available up to the current point in time.

\item Optimization and Trading Strategy:  
The method optimizes three parameters—upper and lower barrier factors (\(F_\text{u}, F_\text{l}\)) and the vertical barrier length (\(\text{V}_t\))—over six-month intervals using the Sharpe Ratio (\(\text{SR}\)) calculated as shown by Eq.~(\ref{eq:sharpe}):  
\begin{align}
\text{SR} = \frac{\mathbb{E}[R_p - R_f]}{\sigma_p}
\label{eq:sharpe}
\end{align}
where \(R_p\) is the portfolio return, \(R_f\) is the risk-free rate (set to 4\% annually), and \(\sigma_p\) is the standard deviation of portfolio returns. A 365-day year is assumed. The optimization ensures that barriers (\(\text{U}_t\), \(\text{L}_t\), and \(\text{V}_t\)) are only valid if touched by the high or low price within \(\text{V}_t\). A minimum trend duration of two days is enforced to enhance labeling robustness. For a bullish window, a long position is opened at the start of the window and held until the window ends. For a bearish window, a short position is opened at the start of the window and closed at the end of the window. No trades are executed during neutral windows.

\end{itemize}

\begin{figure}[t!]
\centering
\includegraphics[width=\linewidth]{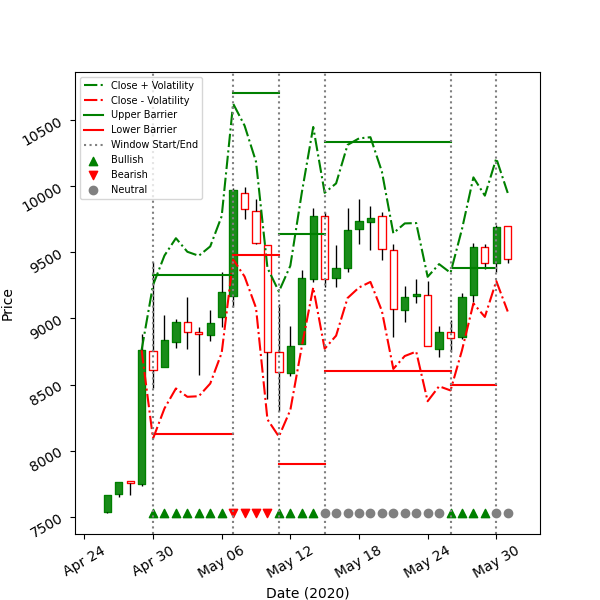}
\caption{Visualization of the proposed labeling method. The green and red lines represent the upper and lower barriers respectively, while the markers denote assigned labels for the given trend windows (denoted with vertical dotted lines).}
\label{fig:tbl_vis}
\end{figure}

This labeling process aligns with the typical time frame of social media driven market impact, providing an effective decision-making window for trading agents. It addresses challenges in early-window predictions while becoming more reliable as the window progresses. Figure~\ref{fig:tbl_vis} illustrates this method, with green and red lines indicating upper and lower barriers, and assigned labels as markers. By directly linking textual inputs and price changes to market movements, this approach overcomes traditional FSA limitations and enhances predictive and actionable utility.

\subsection{Short-Term Trend Classification of Tweets}

We introduce three language model classifiers tailored to predict short-term market trends using our novel market-derived labeling approach. The primary objective is to assess whether language models can directly predict market movements from tweets and how they compare against conventional sentiment-based classifiers. Initially, we evaluate the capability of language models to directly predict market trends by examining whether tweets contain learnable features indicative of subsequent market movements. We start by benchmarking traditional sentiment models, specifically FinBERT and CryptoBERT, against our dataset labeled with market-derived trends. This initial step sets the baseline for understanding if sentiment alone can predict market shifts. Following this, we fine-tune the CryptoBERT model on these market-derived labels, creating what we call the Context-UnAware (CUA) model. This model is designed to detect patterns in tweet content that correlate with market trends, preserving the model's existing knowledge of cryptocurrency jargon and tweet structures. The fine-tuning process involves freezing the first 11 layers of CryptoBERT to maintain this domain-specific knowledge while only adjusting the classification head. The training was conducted over 2 epochs with a 5-fold cross-validation strategy to avoid overfitting, using the AdamW optimizer with a learning rate of $10^{-5}$ and a batch size of 12. A learning rate scheduler with a linear warm-up phase (10\% of steps) was employed to facilitate gradual learning. Building on the insights from previous research on the synergy between market context and textual content \cite{man2019financial}, we then introduce market context into our model's input through prompt engineering. This technique has proven effective in financial forecasting and sentiment analysis \cite{leveraging_ahmed_2024}. We leverage prompt tuning to enhance the predictive power of a pre-trained language model without altering its architecture \cite{lester2021power}. The features added to each tweet's prompt include the ROC, RSI, and the trend of the previous time window, selected due to having the highest correlations with market trends among the available features ($0.26$ for previous window trend, $0.14$ for ROC, and $0.13$ for RSI). These indicators are transformed into descriptive terms, forming prompts like:
\begin{quote}
Previous Label: bullish, ROC: neutral, RSI: bearish, Tweet: Bitcoin is on the rise!
\end{quote}

The Context-Aware (CA) model, is an evolution of the CUA model, fine-tuned using these enriched prompts. Recognizing the importance of temporal context in market analysis, we further refine our model by incorporating the date of each tweet into the prompts. Language models typically do not possess inherent temporal awareness due to their non-sequential processing \cite{timeaware_dhingra_2021, agarwal2022temporal}. By adding this temporal element, we aim to enhance the model's sensitivity to time-specific market conditions, leading to the development of the Temporal Context-Aware (TCA) model. An example of this prompt includes:

\begin{quote}
Date: 2020, January, 01, Previous Label: bullish, ROC: neutral, RSI: bearish, Tweet: Bitcoin is on the rise!
\end{quote}

Through these methodologies, we expand the feature space available to the language model, enabling it to discern complex interrelations between textual content, market indicators, and temporal dynamics, thereby improving predictive accuracy in FSA.

\subsection{Signal Generation and Backtesting Strategies}
The classification of tweets is only the first step in the FSA pipeline. Once tweets are classified, the next task is to aggregate these results over a given timeframe to generate a definitive trading signal. These signals can then be used to take trading positions directly or as input to more complex models in combination with price-based or other signals. However, in this study, we focus on using our model as a single, end-to-end short-term trend classifier, excluding fusion with price-based predictions since our approach already integrates market-related information into the prompts. For comparative purposes, we also evaluate a price-based LSTM model similar to the one proposed by PreBit \cite{prebit_zou_2023}, and a pre-trained time-series masking autoencoder model inspired by \cite{zha2022time, dong2024simmtm} a simple fusion model that combines sentiment and price signals into one. This allows us to compare their performance with the signal generation methods derived from tweet classifications. We propose two methods for signal generation based on tweet aggregation, the majority and mean methods. These methods help generate both the direction and size of each trading position, which are later used in our backtesting experiments. The majority vote method aggregates daily tweet predictions by determining the majority sentiment class for a given day. Let \( N \) denote the total number of tweets in a day, and let \( N_{\text{bullish}} \), \( N_{\text{bearish}} \), and \( N_{\text{neutral}} \) represent the number of tweets classified as \textit{Bullish}, \textit{Bearish}, and neutral, respectively. The majority signal, denoted as \( S_{\text{majority}} \), is defined by Eq.~(\ref{eq:sig_maj}): 
\begin{align}
\label{eq:sig_maj}
S_{\text{majority}} = \arg\max \left( N_{\text{bullish}}, N_{\text{bearish}}, N_{\text{neutral}} \right)
\end{align}

The confidence of the majority class, \( C_{\text{majority}} \), is a measure of how dominant this class is among all classified tweets. It is calculated as given by Eq.~(\ref{eq:conf_maj}) the proportion of tweets belonging to the majority class:
\begin{align}
\label{eq:conf_maj}
C_{\text{majority}} = \frac{N_{\text{majority}}}{N}
\end{align}
Confidence reflects the certainty of the aggregated sentiment signal. Higher confidence indicates that a significant portion of the tweets for the day align with the majority sentiment, whereas lower confidence suggests greater variability in sentiment. This confidence score is used as a factor in determining the size of the trading position. To ensure comparability, the confidence values are normalized on a scale of 0 to 1 using min-max normalization. This step adjusts the confidence scores relative to the range of values observed across the dataset. The mean method aggregates tweet predictions by averaging the encoded class values over a day. In this method, class encodings are defined as follows: \textit{Bearish} = 0, \textit{Neutral} = 1, and \textit{Bullish} = 2. Let \( E_i \) denote the encoded class of the \( i_{th} \) tweet. The mean sentiment for a day, denoted as \( D_{\text{mean}} \), is calculated by Eq.~(\ref{eq:sig_mean}):
\begin{align}
\label{eq:sig_mean}
D_{\text{mean}} = \frac{1}{N} \sum_{i=1}^{N} E_i.
\end{align}

The thresholds \( T_{\text{bearish}} \) and \( T_{\text{bullish}} \) are used to determine the overall daily signal in comparison to \( D_{\text{mean}} \) as given by Eq. \ref{eq:signal_mean}, and the thresholds in this method are optimized every 6 months to match the market preferences:
\begin{align}
\label{eq:signal_mean}
S_{\text{mean}} = 
\begin{cases} 
\text{Bearish} & \text{if } D_{\text{mean}} < T_{\text{bearish}}, \\
\text{Bullish} & \text{if } D_{\text{mean}} > T_{\text{bullish}}, \\
\text{Neutral} & \text{otherwise}.
\end{cases}
\end{align}

\begin{table*}[t]
\small
\caption{Evaluation of base models Base-1, and Base-2, (respectively FinBERT \cite{huang2023finbert} and CryptoBERT \cite{sentiment_kulakowski_2023}) and proposed methods on the 2020 5-fold cross-validation dataset ($E_a$) and event-sampled Bitcoin data ($E_b$). The results for $E_a$ are the averages of metrics across 5 folds, with standard deviations representing fold-to-fold variation. For the base models, $E_a$ results reflect variations across the different data folds.}
\label{tbl:context_aware}
\begin{tblr}{
  cells = {c},
  cell{1}{1} = {r=2}{},
  cell{1}{2} = {c=2}{},
  cell{1}{4} = {c=2}{},
  cell{1}{6} = {c=2}{},
  cell{1}{8} = {c=2}{},
  cell{1}{10} = {c=2}{},
  hline{3} = {1-12}{},
  colsep=1.5mm
}
Model  & Accuracy      &               & Precision     &               & Recall        &               & F1-Score      &               & Cross-entropy Loss &      \\
       & $E_a$         & $E_b$         & $E_a$         & $E_b$         & $E_a$         & $E_b$         & $E_a$         & $E_b$         & $E_a$ & $E_b$ \\
Base-1 & $33.3\pm0.6$ & 32.8 & $33.8\pm2.3$ & 26.4 & $33.3\pm0.1$ & 33.2 & $17.2\pm0.3$ & 16.7 & $2.06\pm5.1$ & 2.41 \\
Base-2 & $33.3\pm1.3$ & 28.2 & $32.2\pm1.9$ & 30.8 & $33.3\pm0.9$ & 32.1 & $27.7\pm0.8$ & 25.2 & $2.28\pm9.1$ & 2.37 \\
CUA    & $44.1\pm1.3$ & $35.3\pm0.5$ & $44.7\pm1.9$ & $34.4\pm0.7$ & $44.1\pm1.3$ & $35.8\pm0.4$ & $43.0\pm2.2$ & $33.3\pm0.8$ & $1.04\pm0.12$ & $1.14\pm0.15$ \\
CA     & \textbf{$89.6\pm9.6$} & \textbf{$81.0\pm0.2$} & \textbf{$90.3\pm12.1$} & \textbf{$80.4\pm0.3$} & \textbf{$89.3\pm14.4$} & \textbf{$80.2\pm0.2$} & \textbf{$89.5\pm13.2$} & \textbf{$80.3\pm0.2$} & \textbf{$0.40\pm0.05$} & $1.24\pm0.11$ \\
TCA    & $86.5\pm8.6$ & $71.2\pm1.3$ & $86.6\pm5.9$ & $71.2\pm1.0$ & $86.5\pm7.9$ & $71.8\pm1.4$ & $86.5\pm6.9$ & $71.0\pm1.4$ & $0.72\pm0.06$ & \textbf{$0.76\pm0.07$}
\end{tblr}
\end{table*}

The confidence score for this method is calculated as the absolute difference between the daily mean \(D_mean\) and the corresponding threshold, reflecting how strongly the mean aligns with a specific sentiment. As with the majority vote method, the confidence scores are normalized to facilitate consistent position sizing across different days. Both methods ultimately provide a daily sentiment signal \( S \) and its corresponding confidence \( C \), which are then used in backtesting to evaluate the performance of the trading strategy. By linking confidence to position size, the approach ensures that trading decisions reflect the model's certainty in its predictions.

We propose three primary strategies for evaluating the effectiveness of our sentiment-driven trading signals. The first strategy, the \textit{TBL} strategy, uses volatility-based dynamic barriers (as defined by Eq.~(\ref{eq:upper}), and Eq.~(\ref{eq:lower})) to define take-profit and stop-loss levels, allowing for adaptive trading decisions based on market conditions. Optimized parameters for barrier thresholds and time limits are tuned for each evaluation period to maximize strategy performance. The second set of strategies referred to as \textit{In-Out} strategies, employ aggregated daily signals to determine entry and exit points. The \textit{In-Out-Long} strategy opens long positions when a \textit{Bullish} signal is detected and closes them upon receiving a \textit{Bearish} signal, while the \textit{In-Out-Short} strategy operates inversely, opening short positions with \textit{Bearish} signals and closing them with \textit{Bullish} ones. These strategies incorporate normalized confidence scores derived from the sentiment classification process to scale position sizes, ensuring that trades reflect the model's certainty in its predictions. Additionally, we use a baseline buy-and-hold strategy for comparison, which provides a simple benchmark to contextualize the performance of the proposed methods. The backtesting process is conducted using the \texttt{vectorbt} library\footnote{github.com/polakowo/vectorbt}, offering a comprehensive and efficient framework for evaluating trading strategies across varying market conditions. Recognizing the data dependency of backtesting results, we evaluate the proposed strategies over multiple market regimes, spanning different periods of bullish, bearish, and neutral trends. This ensures a robust assessment of strategy performance and generalizability. Key performance metrics include profitability measures, such as cumulative return and daily return; risk measures, including maximum drawdown and volatility; and profit-to-risk ratios like the Sharpe and Sortino ratios. By analyzing these metrics, we identify the strengths and weaknesses of each strategy relative to the buy-and-hold baseline, providing insights into the applicability of sentiment-driven trading signals in diverse market environments.

\section{Results}
\label{sec:results}
In this section, we first compare the performance of traditional sentiment models against our proposed approach in classifying tweets on the proposed short-term trend labels. Then, we generate daily trading signals based on the aggregation of daily predictions of the proposed model and compare them to the price-based classification models and conventional fusion models utilizing price and sentiment features. Finally, the proposed trading strategies are evaluated through backtesting across various market conditions.

\subsection{Tweet Classification Evaluation}

\begin{figure*}[t]
    \centering
    \includegraphics[width=\textwidth]{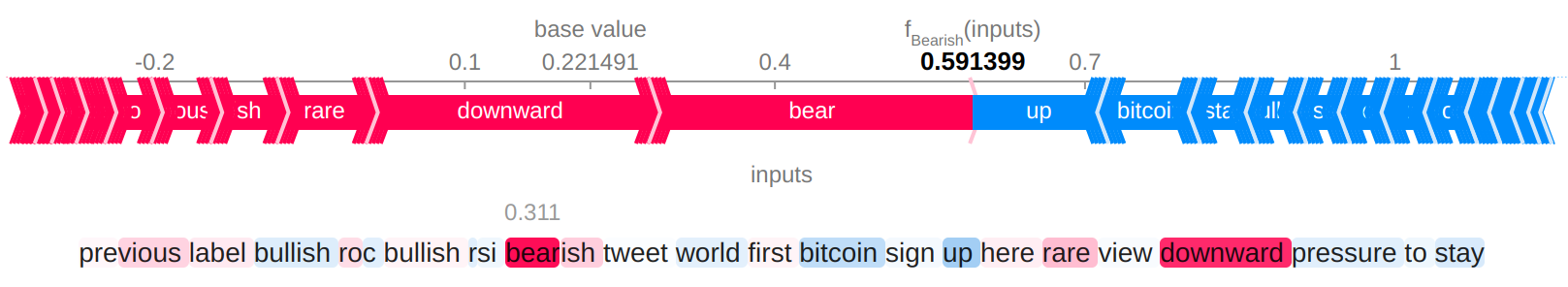}
    \caption{SHAP values plot for a sample tweet. The plot shows the contribution of each word to the model's prediction, highlighting the significant impact of the RSI feature on the \textit{Bearish} prediction.}
    \label{fig:shap}
\end{figure*}

This experiment investigates the ability of language models to predict short-term market trends as obtained by the proposed labeling method. The primary objective is to evaluate whether fine-tuning sentiment models on market-derived labels and incorporating contextual information can enhance the performance of traditional sentiment models in forecasting short-term market movements. The models are fine-tuned on a dataset from 2020 ($E_a$), consisting of approximately 60,000 tweets evenly distributed across the three trend classes, and evaluated on both $E_a$ and an event-sampled dataset ($E_b$). The $E_b$ dataset includes around 40,000 tweets from 2015 to 2023 (excluding 2020) and focuses on tweets related to significant Bitcoin events. This dataset represents high-volatility, news-rich scenarios, providing an opportunity to test the robustness of the models under diverse market conditions. To ensure a rigorous evaluation, grouped 5-fold cross-validation is employed, preventing information leakage by grouping tweets based on their temporal proximity. The training is conducted over 2 epochs per fold, as experiments demonstrate that performance plateaus beyond this point, indicating that the models effectively converge within this limit. Baseline models, FinBERT \cite{huang2023finbert} and CryptoBERT \cite{sentiment_kulakowski_2023}, respectively called \textit{Base-1}, and \textit{Base-2} in Table \ref{tbl:context_aware}, exhibit poor performance in predicting short-term market trends, achieving F1-scores of 17.2\% and 27.7\% on $E_a$, respectively. These results highlight the limitations of sentiment-based models in capturing short-term market movements. This underperformance can be attributed to the fact that these models are trained to interpret sentiment labels based on human annotations, which reflect subjective sentiment rather than actual market behavior. Consequently, the models cannot connect textual features with market-derived outcomes, as they are not explicitly trained on labels derived from market movements. In contrast, the CUA model, fine-tuned with the proposed market-derived labels, demonstrates significant improvement, achieving an F1-score of 43.0\% on $E_a$ highlighting the importance of using market-derived labeling to bridge the gap between textual features and market dynamics.

Further performance gains are achieved through prompt tuning with market context. The CA model, which incorporates features such as the previous trend, ROC, and RSI as textual input, achieves F1-scores of 89.5\% on $E_a$ and 80.3\% on $E_b$. Compared to the CUA model, the CA model demonstrates an impressive improvement across all evaluation metrics, highlighting the pivotal role of descriptive market context in bridging the gap between textual and market data. These results show that when integrated as textual features, market context empowers the language model to establish patterns between market context, tweet content, and ensuing market trends far more effectively. This improvement is not merely incremental but transformative, as it validates the initial hypothesis that language models when equipped with descriptive market context, can function as end-to-end predictors for short-term trends without the need for fusion with price-based models. 

As shown in Table \ref{tbl:context_aware}, the TCA model, which incorporates temporal context alongside market features, demonstrates notable performance on the 2020 dataset ($E_a$), with an F1-score of 86.5\% (±6.9\%), and on the event-sampled dataset ($E_b$), achieving an F1-score of 71.0\% (±1.4\%). While the model performs well on $E_a$, its performance on $E_b$ is noticeably lower, suggesting that the inclusion of temporal context may increase the risk of overfitting to the training data. This reduced performance on the unseen event-sampled data indicates that the model may not generalize as effectively as the CA model, which performs better on $E_b$ with an F1-score of 80.3\% (±0.2\%). The CA model also performs well on $E_a$ with an F1-score of 89.5\% (±13.2\%) and on $E_b$ with a score of 80.3\% (±0.2\%), but similarly exhibits considerable variability. This variability in performance across folds is more pronounced in the CA and TCA models compared to the base models, where performance remains more stable. These variations highlight the sensitivity of the models to specific data splits. To mitigate these fluctuations, we adopt a majority voting strategy for the subsequent experiments, where predictions are aggregated across all five folds. This ensemble learning approach reduces the influence of fold-specific anomalies, enhances stability, and provides a more reliable evaluation of model performance. Additionally, paired t-tests comparing the CA and CUA models against Base-1 (CryptoBERT) model across the five folds confirm statistically significant differences in performance ($p < 0.001$ for both comparisons), further validating the robustness and superior performance of the proposed models.

To further investigate the contribution of individual context components (previous label, ROC, and RSI), we systematically exclude each feature from the prompts and measure the resulting accuracy. Excluding the previous label causes the model’s accuracy to drop significantly to 43.6\%, highlighting its pivotal role in aligning tweet content with past market trends. On the other hand, removing RSI and ROC results in accuracies of 79.0\% and 78.8\%, respectively, indicating that these features, while less critical than the previous label, still contribute substantially to model performance. To confirm that the model is not overly reliant on prompts and effectively interprets tweet content, we conducted extensive SHAP value analysis \cite{unified_lundberg_2017} on a diverse range of tweets. Our tests revealed a wide variety of contributions from different parts of the tweets, demonstrating the model's ability to adapt to varying linguistic patterns and contexts. Figure~\ref{fig:shap} illustrates one such sample from our analysis, showcasing how specific words influence the model’s prediction. In this example, despite a \textit{Bullish} previous label, the model predicts a \textit{Bearish} outcome primarily due to the strong contribution of the word ``downward'' in the tweet. The SHAP values indicate that ``downward'' exerts a dominant influence on the prediction, strongly shifting the model’s decision toward bearishness. This analysis confirms that the model effectively processes tweet content, ensuring predictions are based on meaningful contextual analysis rather than superficial reliance on the prompt information. The results validate the robustness of the model in integrating contextual information and analyzing tweet content for accurate market predictions.

\subsection{Signal Generation}
\begin{table}[t]
\centering
\small
\caption{Evaluation of benchmark price-based models (Base-3, and Base-4), and fusion models, alongside majority and mean signal generation methods based on the proposed language-based prediction model (CA).}
\label{tbl:next_day}
\begin{tblr}{
  row{odd} = {c},
  row{4} = {c},
  row{6} = {c},
  row{8} = {c},
  row{10} = {c},
  row{12} = {c},
  cell{1}{1} = {c=2}{},
  cell{2}{1} = {r=5}{},
  cell{2}{2} = {c},
  cell{2}{3} = {c},
  cell{2}{4} = {c},
  cell{2}{5} = {c},
  cell{2}{6} = {c},
  cell{7}{1} = {r=3}{},
  cell{10}{1} = {r=3}{},
  hline{2,7,10} = {-}{},
}
Model                               &                     & Accuracy       & Precision      & Recall         & F1             \\
\begin{sideways}OVR\end{sideways}   & 

                                    Base-3 \cite{prebit_zou_2023}         & 52.87          & 46.76          & 52.87          & 49.20          \\
                                    & Base-4 \cite{zha2022time, dong2024simmtm} & 54.25          & 51.31          & 54.63          & 50.01          \\
                                    & Fusion              & \textbf{59.01} & 58.35 & 55.14          & \textbf{55.94} \\
                                    & Majority (CA) & 53.52          & \textbf{61.15}          & \textbf{57.46} & 54.21          \\
                                    & Mean (CA)    & 57.91          & 56.50          & 57.13          & 56.10          \\
\begin{sideways}OVO +\end{sideways} & Fusion              & 78.38          & 80.72          & \textbf{90.38} & 85.28 \\
                                    & Majority (CA) & \textbf{84.57} & \textbf{84.30} & 88.02 & \textbf{86.12}          \\
                                    & Mean (CA)    & 76.95          & 83.60          & 76.33          & 79.80          \\
\begin{sideways}OVO -\end{sideways} & Fusion              & 78.38          & 70.31          & 51.33          & 59.34          \\
                                    & Majority (CA) & \textbf{84.92} & \textbf{84.92} & \textbf{80.45} & \textbf{82.62} \\
                                    & Mean (CA)     & 76.95          & 69.00          & 77.86          & 73.16          
\end{tblr}
\end{table}

The signal generation process translates tweet classifications into actionable trading signals, providing a clear recommendation on the short-term market movement. Unlike the classification task, which focuses on individual tweets, this step aggregates predictions from all tweets in a given day to determine the dominant signal. Traditional methods for signal generation often rely on multi-stage architectures, where sentiment predictions are combined with market-derived features in a separate fusion model to return a final prediction of how the market will move. While effective in some cases, such architectures fail to capture nuanced interactions between textual and market features, leading to potential information loss and requiring more data and a more complex architecture. In contrast, our CA language model integrates market features (e.g., previous trend, ROC, RSI) directly into tweet prompts as textual descriptions, allowing for an end-to-end prediction process that inherently models these interactions. This experiment evaluates whether this innovative approach generates reliable signals and how it compares to traditional price-based and fusion-based prediction models.

For comparative analysis, we utilize two baseline price-based models, Base-3 (LSTM model inspired by \cite{prebit_zou_2023}) and Base-4 (Autoencoder model inspired by \cite{zha2022time, dong2024simmtm}), along with a fusion model that combines predictions from the base CryptoBERT model, price-based models, and our proposed majority CA model to generate short-term market signals. For each tweet, our proposed model generates predictions using an ensemble majority voting method across a 5-fold cross-validation setup, ensuring data diversity among the models. These individual predictions are then aggregated daily using either the majority or mean aggregation methods. The fusion model takes the daily predictions based on the majority aggregation method as input, alongside traditional sentiment and price-based predictions, to ensure it has access to all available features. The final output from the model is a predicted class (bullish, bearish, or neutral) for each day. For sentiment features, the fusion model receives the ratio of bullish to bearish tweets, represented as a continuous value between 0 and 1, which has proven to be more effective than using the final prediction class. Similarly, price-based models provide a final predicted class utilizing features such as OHLCV, Ethereum price, gold price, RSI, ROC, MACD, and moving averages (SMA7, SMA21, EMA12, EMA26). These three input features (the CA model's aggregated predictions, the sentiment ratio, and the price-based predictions) are then combined in the fusion model to generate the final market prediction, which is used to assess short-term market movements.

\begin{figure}[ht]
\centering 
\includegraphics[width=1\linewidth]{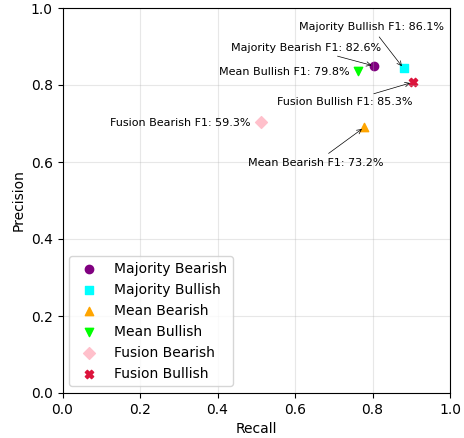} \caption{Precision-recall scatter plot comparing performance across different methods for \textit{Bullish} and \textit{Bearish} predictions. The majority and mean aggregation methods and the fusion method demonstrate their precision, recall, and F1-scores.} 
\label{fig:true_p}
\end{figure}

In Table \ref{tbl:next_day}, Metrics accuracy, precision, recall, and F1-score are presented in both One-Vs-Rest (OVR) and One-Vs-One (OVO) modes for our benchmarks and proposed language-based models. The OVR mode assessed the models’ ability to distinguish between \textit{Bullish}, \textit{Bearish}, and \textit{Neutral} predictions. In contrast, the OVO mode focused on pairwise comparisons between \textit{Bullish} and \textit{Bearish} predictions (presented with \textit{OVO +} for bullish against bearish, and \textit{OVO -} for bearish against bullish metrics), which have a greater trading impact. This approach ensured a comprehensive evaluation of the models' capabilities, with the OVO results highlighting sensitivity to critical decision-making scenarios. The results, presented in Table~\ref{tbl:next_day}, indicate that the CA model achieved comparable or superior performance to the fusion and price-based models. For example, the majority vote aggregation of the CA model attained an F1-score of 54.21\%, outperforming Base-3 model (49.20\%) and Base-4 model (50.01\%). Notably, the majority method demonstrated balanced performance across \textit{Bullish} and \textit{Bearish} predictions, achieving F1-scores of 82.62\% for \textit{Bearish}, and 86.12\% for \textit{Bullish} predictions and consistent precision and recall across these scenarios. In contrast, while the fusion model performed slightly worse in \textit{Bullish} scenarios (F1: 85.28\%), it struggled significantly with \textit{Bearish} predictions (F1: 59.34\%).

\begin{figure*}[t]
    \centering
    \includegraphics[width=1\linewidth]{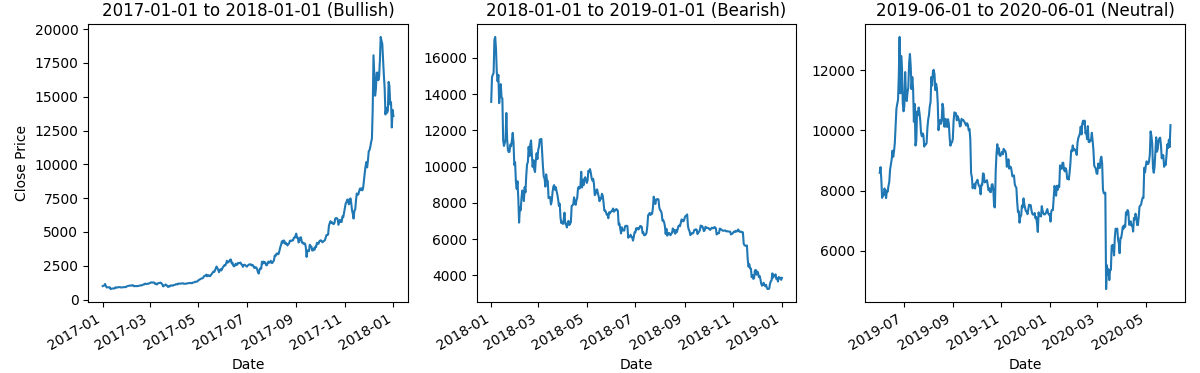}
    \caption{Yearly periods selected to represent three distinct market regimes: \textit{Bullish}, \textit{Bearish}, and \textit{Neutral}.}
    \label{fig:regimes}
\end{figure*}

These findings underscore the adaptability of the proposed method in dynamic market environments and its ability to serve as a reliable end-to-end prediction model. By aggregating tweet classifications with market context descriptors, the proposed method bypasses the need for separate fusion architectures and demonstrates the potential of language models to capture complex interactions between textual and market data. This opens the door to leveraging language models for signal generation in financial forecasting, reducing reliance on multi-stage architectures and offering a streamlined, effective alternative. An important observation from the confusion matrix of the proposed majority method is the nature of misclassifications. Only 5.1\% of \textit{Bullish} labels are misclassified as \textit{Bearish}, and 12.5\% of \textit{Bearish} labels are misclassified as \textit{Bullish}. This demonstrates that the model avoids direct, high-impact errors between \textit{Bullish} and \textit{Bearish} labels—errors that can lead to substantial trading risks. Instead, most misclassifications occur when \textit{Bullish} or \textit{Bearish} labels are predicted as \textit{Neutral}. This behavior helps mitigate risk by avoiding overly aggressive trading signals in uncertain scenarios, highlighting the model's conservative tendencies when dealing with ambiguous market conditions. Figure~\ref{fig:true_p} illustrates a precision-recall scatter plot for \textit{Bullish} and \textit{Bearish} labels, showing the trade-offs achieved by the proposed methods in comparison to the fusion model. This visualization is particularly important, as precision and recall are critical for generating reliable trading signals. The majority method maintains high precision and recall while minimizing risky misclassifications highlighting its robustness and adaptability.

\subsection{Trading Strategy}

\begin{figure}[t]
    \centering
    \includegraphics[width=0.8\linewidth]{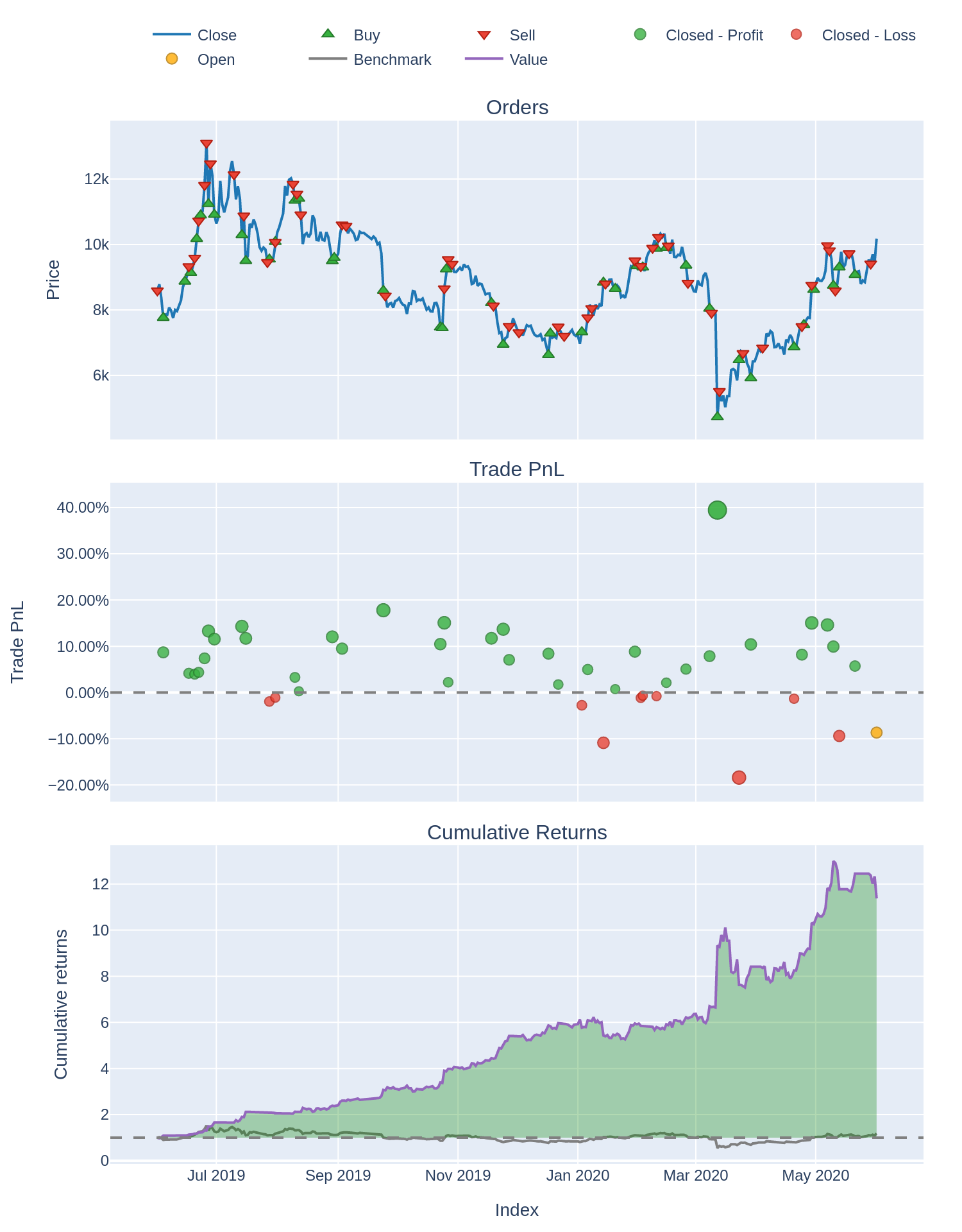}
    \caption{Backtest visualization of the \textit{Mean TBL} strategy during the neutral period, illustrating entry and exit points, as well as the profit and loss for each trade. The chart highlights the underperformance of the buy-and-hold strategy in stagnant market conditions, while demonstrating the effectiveness of our proposed method in navigating such environments.}
    \label{fig:backtest}
\end{figure}
\begin{table*}[ht]
\centering
\small
\caption{Backtest evaluation of our proposed language-model-based approach to signal generation in Bullish, Bearish, and Neutral market regimes.}
\label{tbl:backtest_results}
\begin{tblr}{
  cells = {c},
  cell{1}{2} = {c=2}{},
  cell{2}{1} = {r=7}{},
  cell{2}{2} = {c=2}{},
  cell{3}{2} = {r=3}{},
  cell{6}{2} = {r=3}{},
  cell{9}{1} = {r=7}{},
  cell{9}{2} = {c=2}{},
  cell{10}{2} = {r=3}{},
  cell{13}{2} = {r=3}{},
  cell{16}{1} = {r=7}{},
  cell{16}{2} = {c=2}{},
  cell{17}{2} = {r=3}{},
  cell{20}{2} = {r=3}{},
  hline{2,9,16,23} = {-}{},
  hline{3,10,17, 6,13,20} = {dotted},
}
Regime  & Strategy                               &          & {Daily \\Return \%} & Sharpe~       & Sortino       & {Closed\\Trades} & {Max\\DD\%}      & {Profit \\Factor}       & {Win\\Rate\%} & {Avg Win\\~Trade Time~} & {Avg Losing~\\Trade Time~} \\
Bullish & Buy and Hold                           &          & 3.38       & 3.12          & 4.52          & 1                & 35.36          & -                       & 100            & 367 days                                & -                               \\
        & \begin{sideways}Majority\end{sideways} & In-Out-Long & \textbf{10.88}                & 4.84          & 7.15          & 11               & 34.46          & 807.90          & 90.90            & 23 days                                & 7days                               \\
        &                                        & In-Out-Short & 0.40                & 2.39          & 2.49          & 11               & \textbf{11.11}        & 19.44                   & 72.72            & 6 days                                & 8 days                               \\
        &                                        & TBL      & 7.86                & \textbf{5.07}          & \textbf{8.42}          & 66              & 20.92          & 12.04                    & 84.84            & \textbf{2 days}                                & \textbf{3 days}                               \\
        & \begin{sideways}Mean\end{sideways}     & In-Out-Long & 6.45                & 4.58          & 5.99          & 17               & 21.56          & 290.19                   & 93.75            & 12 days                                & 6 days                               \\
        &                                        & In-Out-Short & 0.01                & 0.36          & 0.37          & 16               & 61.01        & 1.04                   & 68.75            & 7 days                                & 18 days                               \\
        &                                        & TBL      & 4.27                & 3.69 & 6.12 & \textbf{79}     & 47.04          & 2.51                   & 77.21            & \textbf{2 days}                                & 5 days                               \\
Neutral & Buy and Hold                           &          & 0.04       & 0.58          & 0.69          & 1                & 63.97          & -                       & 100            & 367 days                                & -                               \\
        & \begin{sideways}Majority\end{sideways} & In-Out-Long & 0.55                & 3.35          & 6.03          & 6               & 8.09          & -          & 100            & 22 days                                & -                               \\
        &                                        & In-Out-Short & 0.29                & 1.45          & 1.99          & 6               & 22.48      & 28.17                   & 83.33            & 39 days                                & 14 days                               \\
        &                                        & TBL      & 2.22                & 3.70          & 5.29          & 33              & 25.05          & 4.23                    & 90.90            & \textbf{4 days}                                & \textbf{4 days}                               \\
        & \begin{sideways}Mean\end{sideways}     & In-Out-Long & 0.62                & 3.60          & \textbf{7.32}          & 8               & \textbf{6.62}      & -                   & 100            & 9 days                                & -                               \\
        &                                        & In-Out-Short & 0.29                & 1.41          & 2.18          & 8               & 22.94       & 5.62                   & 62.5            & 47 days                                & 11 days                               \\
        &                                        & TBL      & \textbf{2.81}                & \textbf{3.73} & 6.17 & \textbf{45}     & 25.67          & 3.80                   & 77.77            & 5 days                                & 5 days                               \\
Bearish & Sell and Hold                           &          & 0.19       & 1.26          & 1.77          & 1                & 26.45          & -                       & 100            & 367 days                                & -                               \\
        & \begin{sideways}Majority\end{sideways} & In-Out-Long & 0.43                & 1.75          & 	1.92          & 15               & 49.05          & 10.37            & 73.33          & 11 days            & 16 days
        \\
        &                                        & In-Out-Short & 1.23                & 4.21          & \textbf{8.48}          & 15               & \textbf{9.89}  & 69.69                   & 93.33            & 10 days                                & 7 days                               \\
        &                                        & TBL      & \textbf{3.72}                & \textbf{4.70}          & 7.62          & 68              & 33.77          & 7.47                    & 72.05            & \textbf{2 days}                                & \textbf{3 days}                               \\
        & \begin{sideways}Mean\end{sideways}     & In-Out-Long & 0.38                & 1.85          & 1.72          & 17               & 36.01          & 4.86                   & 70.58            & 8 days                                & 8 days                               \\
        &                                        & In-Out-Short & 1.09                & 3.24          & 4.37          & 16               & 32.44  & 17.07                   & 93.75            & 13 days                                & 14 days                               \\
        &                                        & TBL      & 2.30                & 3.67 & 5.60 & \textbf{84}     & 37.68          & 3.08                   & 65.47            & 3 days                                & \textbf{3 days}
\end{tblr}
\end{table*}

After generating signals, we aim to evaluate the performance of our proposed methods in terms of their ability to profit from the market through a few simple trading strategies. A significant challenge in evaluating trading strategies is their dependence on the prevailing market regime, which can influence performance. To ensure a comprehensive assessment, we test our strategies across three distinct market regimes, as shown in Figure~\ref{fig:regimes}. These regimes include a \textit{Bullish} regime from 2017 to 2018, characterized by rapid market growth of approximately 1200\%, a \textit{Bearish} regime from 2018 to 2019, marked by a sharp decline of 250\%, and a \textit{Neutral} regime from mid-2019 to mid-2020, which demonstrated moderate stability with around a 40\% increase. The proposed signal generation methods are applied across three trading strategies. To contextualize these methods, we compare their performance against the \textit{buy-and-hold} and \textit{sell-and-hold} benchmarks, which offer a baseline representing passive strategies that follow market trends. By testing these strategies across diverse market conditions, we aim to understand the adaptability and robustness of the proposed methods in generating actionable trading signals. The results presented in Table \ref{tbl:backtest_results} highlight several notable patterns. During the \textit{Bullish} period, the benchmark buy-and-hold strategy delivers an impressive Sharpe ratio of 3.14 and a daily return of 3.38\%, reflecting Bitcoin's remarkable upward momentum. Despite the benchmark's strong performance, four out of six proposed strategies exceeded this Sharpe ratio, with all strategies outperforming the benchmarks across the other two market regimes, demonstrating robust adaptability. Notably, the \textit{Majority TBL} strategy achieved the highest Sharpe ratio in both the \textit{Bullish} and \textit{Bearish} markets, with a Sharpe ratio of 5.07 in the former and 4.70 in the latter. But, As shown in Figure~\ref{fig:backtest} in the stagnant market regime, the \textit{Mean TBL} method outperformed the others with a Sharpe ratio of 3.73, creating a significant gap over the benchmark. Profiting in stagnant markets is notoriously challenging, but the cumulative return plot in Figure~\ref{fig:backtest} demonstrates a consistent and steady upward trend, showing the robust profitability of our proposed method. As for the \textit{In-Out} strategies, they exhibit a strong dependence on market regimes, as expected given their one-sided trading approach. For instance, the \textit{In-Out-Long} strategy excelled during the bullish regime, delivering high daily returns, but underperformed in bearish conditions. These strategies had longer trade durations than \textit{TBL} strategies, increasing risk exposure, and executed fewer trades, providing respectable Sharpe ratios—except when operating against market trends. Among the proposed methods, the \textit{TBL} strategies emerged as the top performers, with shorter trade durations (typically 2–5 days), smaller maximum drawdowns, and reduced risk exposure. They execute more frequent trades than the \textit{In-Out} strategies and consistently deliver higher Sharpe and Sortino ratios, demonstrating stable performance across different regimes, including profitability in the stagnant neutral markets. Regarding aggregation methods, while the majority and mean approaches deliver comparable overall results, the majority method excelled in trending markets (bullish or bearish), whereas the mean method shows better performance in neutral markets. Finally, testing the \textit{Majority TBL} and \textit{Mean TBL} strategies in January 2021—the first month after the training period and a bullish phase; yielded Sharpe ratios of 4.43 and 3.49, respectively, aligning well with their expected performance in such conditions. 

These results underscore the practicality and robustness of our language model-driven predictions in backtesting environments. Notably, the fine-tuning process, designed specifically with TBL market-derived labels, naturally aligns the model's strengths with the \textit{TBL} strategy. This demonstrates that the choice of labeling method can serve as a strategic advantage, enabling the use of strategies that directly match the labeling process to achieve superior results. The \textit{TBL} strategy stands out as a reliable approach for leveraging model predictions, offering a balanced risk profile and adaptability to a diverse set of market conditions.

\section{Discussion}
\label{sec:discussion}
This section examines the limitations and implications of the proposed approach. We also highlight the benefits of market-derived labeling, context-rich prompts, and language models' potential to enhance financial market predictions and decision-making based on the findings of this study.

\subsection{Limitations}
Our approach assumes a direct short-term market impact for each tweet, which may oversimplify the intricate and often indirect relationship between social media activity and market movements. By treating tweets as market movement descriptors rather than delving into their underlying intent, this method risks overlooking the nuanced factors that drive market behavior. Furthermore, the association between specific textual patterns and market trends may be transient, requiring frequent model updates to maintain performance. The reliance on prompt engineering introduces the risk of overfitting, as the model might become overly attuned to training data patterns, thereby limiting its ability to generalize to new data. While SHAP analysis indicates effective use of tweet content, reducing numerical features to categorical prompts may lose some of the fine-grained details critical for identifying complex patterns, leaving traditional numerical models potentially superior in certain contexts. Despite these challenges, our experiments demonstrate that directly labeling tweets with market outcomes significantly enhances predictive power compared to traditional sentiment-based approaches, offering a more pragmatic framework that prioritizes measurable market impacts in financial social media analysis.

\subsection{Implications}
This study provides insights into the application of sentiment analysis for financial market forecasting, underscoring the need for more context-aware and market-focused approaches. The results highlight the effectiveness of tailoring labeling methods to align with the specific objectives of FSA, emphasizing that predicting market movements rather than merely assessing sentiment, requires a market-derived labeling approach. This finding aligns with previous research \cite{cryptocurrency_lamon_2017} and reinforces the paradigm shift toward directly linking textual content to market behavior \cite{du2024financial}, bypassing the limitations of intermediate sentiment evaluations. Integrating market context through prompt tuning emerges as another pivotal factor in enhancing predictive performance. By embedding key market metrics such as price trends and volatility indicators directly into text prompts, the model captures a richer understanding of the conditions shaping market movements. This approach not only improves prediction accuracy but also demonstrates the potential for augmenting language models with external data sources to achieve even greater efficacy in financial forecasting tasks. Furthermore, the ability of the proposed model to generate actionable trading signals demonstrates its practical utility in real-world financial decision-making. Traders and investors could use these models to inform strategies based on real-time social media analysis, paving the way for more efficient trading practices and enhanced risk management.

\section{Conclusions and Future Work}
\label{sec:conclusion}
In this study, we demonstrate that a direct, market-based approach to classifying tweets outperforms traditional sentiment-based models in predicting Bitcoin’s short-term market movement. Traditional sentiment analysis often fails to bridge the gap between textual sentiment and actual market behavior, as evidenced by the poor performance of the CryptoBERT \cite{sentiment_kulakowski_2023} and FinBERT \cite{huang2023finbert} models, which achieved F1-scores of only 25.2\% and 16.7\%, respectively, on a dataset of historical Bitcoin events. In contrast, by using market-derived labeling—defining clear profit/loss and temporal boundaries—we establish a direct link between text and market outcomes. Fine-tuning language models with these market-derived labels improved the F1-score by 15.3\%, and further integrating market context through prompt-tuning boosted performance by a staggering 47\%, reaching an F1-score of 80.3\%. We also show that our model, which aggregates the predictions of the context-aware language model into actionable signals, can act as an effective standalone short-term market signal generator. Compared to the traditional fusion model approach, our method consistently outperforms benchmarks, achieving Sharpe ratios of up to 5.07 in trending markets and 3.73 in neutral conditions. These results highlight the untapped potential of language-based models for financial forecasting. These findings demonstrate the transformative potential of language models in financial forecasting. These models can produce actionable insights, generate effective trading strategies, and manage risk in highly volatile markets by moving beyond sentiment classification to integrate context and market signals. 

We recommend an in-depth exploration of language-based models for financial decision-making, moving beyond their integration as mere features within traditional numerical forecasting models. Future research could focus on three key areas: First, leveraging LLMs to power chatbots for investment advisory services, assisting financially inexperienced users in navigating investment strategies, best practices, and financial knowledge. Second, advancements in Retrieval-Augmented Generation (RAG) can enable an interactive, query-based interface for financial reports, offering faster, more intuitive access to critical asset information and potentially democratizing detailed financial analysis. Third, improvements in prompt tuning, along with novel feature engineering—such as incorporating tweet author reliability, social media engagement, and technical indicators—could substantially elevate the performance of these models. Additionally, personalizing financial advice by tailoring recommendations to individual profiles, risk tolerance, and goals could transform how financial planning is approached. While the potential of language models in finance is significant, their value must be substantiated through robust evaluation frameworks. These frameworks must simulate real-world conditions, incorporating transaction costs, varying market regimes, and other practical considerations, to fully assess the models' real-world applicability and effectiveness in financial decision-making.

\normalsize
\bibliography{main}
\end{document}